\documentclass[twocolumn,aps,prb,showpacs,superscriptaddress]{revtex4}
\usepackage{graphics,bm}
\usepackage{amssymb}
\usepackage{epsfig}
\usepackage{epsf}

\def\Im{{\rm Im}}
\def\be{\begin{equation}} \def\ee{\end{equation}}
\def\bea{\begin{eqnarray}} \def\eea{\end{eqnarray}}

\def\nn{\nonumber}

\begin{document}

\title{Superconductivity-induced changes in density-density correlation function enabled by Umklapp processes}

\author{Wei-Cheng Lee}
\email{wlee@binghamton.edu}
\affiliation{Department of Physics, Applied Physics, and Astronomy, Binghamton University - State University of New York, Binghamton, USA}

\date{\today}

\begin{abstract}
Motivated by the mid-infrared scenario for high-temperature superconductivity proposed by Leggett, the effects of Umklapp processes on the density-density correlation function 
in the presence of long-range Coulomb interaction 
have been investigated on a microscopic model. We show that because Umklapp processes enable scatterings 
that conserve total momentum only up to $n\vec{K}$, where $n$ is an integer and $\vec{K}$ is the reciprocal wave vector, significant amounts of spectral weight in the 
plasmonic excitations at long wavelength are transferred into lower frequency around the midinfrared regime. We further find that 
regardless of the gap symmetry, superconductivity generally suppresses the Umklapp scatterings due to the formation of the electron pairing. 
This suppression is unique for the superconductivity due to the interplay between electron pairing and the odd parity of the matrix elements associated with Umklapp channels, which usually does not occur
in other known competing orders. Specific predictions for the experimental signatures in optical conductivity and electron energy loss spectroscopy will be discussed.
\end{abstract}

\maketitle

\section{Introduction}
Dynamical responses to experimental probes from the high-temperature superconductors are an important subject in condensed matter physics, since they may hold key information 
to resolve the origin of the superconductivity\cite{dagottormp1994,leermp2006,armitagermp2010,stewartrmp2011,scalapinormp2012}. While enormous efforts have been made to understand the novel behaviors 
appearing at low energy ($\sim 100$ meV), including but not limited to, angle resolved photoemission spectroscopy (ARPES)\cite{arpes2003,arpes2012}, inelastic neutron scattering\cite{neutron2012}, 
point contact spectroscopy\cite{arham2012,arham2013,leewcpnas}, and quasiparticle interferences (QPI)\cite{qpi2011}, the study of dynamical responses 
at a frequency in the midinfrared regime or even higher has received much less attentions. Perhaps it is due to the belief that although the physics at high energy may play a role 
in the pairing mechanism at low energy, it is unlikely affected by the superconductivity, 
given that the superconducting gap is only around the order of 10-50 meV.

In this regard, the observation of significant changes in optical conductivity due to the superconductivity up to 
energy as high as the order of 1 eV in cuprates\cite{molegraaf2002,boris2004,kuzmenko2005,carbone2006} and iron-based superconductors\cite{boris2011} 
is a shocking result. These experiments found significant differences in the optical conductivity between the normal and superconducting states in a wide range of frequencies.
According to the BCS theory, regardless of the origin of the pairing mechanism, only the 
electronic structure at energy scales comparable to the superconducting gap is strongly modified. As a result, the dynamical responses should remain unchanged at frequency much higher than the 
superconducting gap, as shown in the seminal papers of Anderson on the theory of plasmon excitation in superconductors\cite{anderson1958,anderson19582}.
Therefore, these unusual changes at high energy are naively attributed to the strong local interaction such as Hubbard $U$ or Heisenberg $J$, but a detailed theory is still lacking.

A even more fundamental question raised by Turlakov and Leggett\cite{leggettpnas1999,turlakov2003} places more constraints on the study of the superconductivity-induced changes in the optical
conductivity. Based on a rigorous consideration of the three sum rules
\be
J_{n=-1,1,3}(\vec{q}) = \frac{2}{\pi}\int_0^\infty d\omega \omega^n \Im \chi(\vec{q},\omega),
\ee
where $\chi(\vec{q},\omega)$ is the density-density correlation function, they found a general statement about the upper and the lower bounds on the Coulomb energy at long 
wavelength by applying Cauchy-Schwartz inequalities. Moreover, it is shown that without processes breaking the conservation of the total momentum of electrons, 
no observable change in the density-density correlation function at long wavelength shall be allowed.
Since the optical conductivity is closely related to the density-density correlation function at long wavelength, the same conclusion applies to optical conductivity as well.
In other words, the strong local interactions such as $U$ and $J$ {\it alone} can not explain the changes in optical conductivity, and the interaction breaking the conservation of total momentum 
of electrons has to be identified to understand this experimental puzzle. 

Since all the high-temperature superconductors known to date are crystalline and have Fermi surfaces close to the Brillouin-zone boundary, Umklapp processes are the most ubiquitous 
momentum-conservation-breaking terms. While in the semiconductors the Umklapp processes have been included in the first-principle calculations, known as the
local field effect\cite{hanke1974,hanke1975}, the role of Umklapp processes in the correlated materials as well as in the superconducting states is still not understood.
In this paper, we investigate the effects due to Umklapp processes on the density-density correlation function in both normal and superconducting states.
We find that a significant amount of spectral weight is created at frequencies below the plasmon frequency due to the presence of Umklapp processes.
In the superconducting state, the interplay between the nature of electron pairing between $(\vec{k}\uparrow)$ and $(-\vec{k}\downarrow)$  and the odd parity of the matrix elements associated
with Umklapp processes substantially suppresses the effects from Umklapp processes.
This superconductivity-induced suppression of Umklapp processes results in the decrease of the spectral weights in the frequency range well above the gap but below
the plasmon frequency, and should occur as the system undergoes the superconducting phase transition, consistent with the changes of optical conductivity observed experimentally.
Moreover, we predict that a downward shift of plasmon frequency as well as an increase of the spectral weight in the plasmon modes should occur simultaneously.
We will show that these signatures could be revealed from the analyses of the existing data of Ref. [\onlinecite{carbone2006}] and
recent measurement by Levallois {\it et. al.}\cite{julien} on the optimally doped Bi$_2$Sr$_2$Ca$_2$Cu$_3$O$_{10}$.
Further experiments on optical conductivity and electron energy loss spectroscopy (EELS) will be necessary for future study.

\section{Hamiltonian of a two-dimenstional system with periodic potential along $\hat{x}$ direction}
We start from the general Hamiltonian with a periodic potential along the $\hat{x}$ direction
\bea
H&=&H_K + H_{Coul}\nn\\
H_K&=&\int d\vec{r} \psi^\dagger_{\vec{r}\sigma}\big[-\frac{\hbar^2\nabla^2}{2m} - \mu + 2U\cos(\vec{K}_x\cdot\vec{r}) \big] \psi_{\vec{r}\sigma},\nn\\
H_{Coul}&=& \frac{1}{2\Omega} \sum_{\vec{q}\neq 0} v_q\big[\hat{\rho}(\vec{q})\hat{\rho}(-\vec{q})-\hat{N}\big],
\label{hamil}
\eea
where $v_q =e^2/2\epsilon_0\epsilon_\infty q$ for two dimensions (2D) and $\vec{K}_x=\frac{2\pi}{a}\big(\hat{x},0\big)$. 
$\hat{\rho}(\vec{q})$ and $\hat{N}$ are the density and total electron number operators, respectively.
Performing the Fourier transformation on $H_K$, we have
\bea
&&H_K=\sum_{l=-\infty}^{\infty(odd)} \sum_{\vec{k}\sigma}
\epsilon(\vec{k}+l\vec{K}_x/2) c^\dagger_{\vec{k}+l\vec{K}_x/2,\sigma}
c_{\vec{k}+l\vec{K}_x/2\sigma}\nn\\
 &+& U\big[c^\dagger_{\vec{k}+l\vec{K}_x/2\sigma}c_{\vec{k}+(l-2)\vec{K}_x/2\sigma} + h.c.\big],\nn\\
\label{hk}
\eea
where we have introduced a shorthand notation for the integration over momentum as:
\be
\sum_{\vec{k}\sigma}\equiv \sum_\sigma\frac{1}{4\pi^2}\int_{-\frac{K_x}{2}}^{\frac{K_x}{2}} dk_x \int_{-\infty}^\infty dk_y
\ee
Moreover, $\epsilon(\vec{k})\equiv\frac{\hbar^2 k^2}{2m}-\mu$, and $c_{\vec{p},\sigma}$ is the Fourier component of $\psi_{\vec{r}\sigma}$ defined as
\be
\psi_{\vec{r}\sigma} = \frac{1}{4\pi^2}\int dp_x dp_y e^{i\vec{p}\cdot \vec{r}} c_{\vec{p},\sigma}.
\label{plane}
\ee
 
The simplest case is to consider only two $l$'s, which we pick as $l=\pm 1$. This choice satisfies all of the necessary symmetries including time reversal, parity, etc., and therefore 
it serves as an excellent example for the proof of principles. We can then reduce the $H_K$ to

\bea
H_K&=&\sum_{\vec{k}\sigma} \epsilon(\vec{k}) c^\dagger_{\vec{k}-\vec{K}_x/2,\sigma}c_{\vec{k}-\vec{K}_x/2,\sigma} + c^\dagger_{\vec{k}+\vec{K}_x/2,\sigma}c_{\vec{k}+\vec{K}_x/2,\sigma}\nn\\
&+&U\big[c^\dagger_{\vec{k}-\vec{K}_x/2\sigma}c_{\vec{k}+\vec{K}_x/2\sigma}
+ c^\dagger_{\vec{k}+\vec{K}_x/2\sigma}c_{\vec{k}-\vec{K}_x/2\sigma}\big],
\eea
which can be diagonalized as
\be
H_K=\sum_{\vec{k}\sigma} E^+_{k\sigma}c^\dagger_{+,\vec{k}\sigma}c_{+,\vec{k}\sigma} + E^-_{k\sigma}c^\dagger_{-,\vec{k}\sigma}c_{-,\vec{k}\sigma}
\ee
where 
\bea
E^\pm_{k\sigma}&\equiv& \epsilon_1(\vec{k}) \pm D(\vec{k})\nn\\
\epsilon_1(\vec{k})&\equiv&\frac{\epsilon(\vec{k}-\vec{K}_x/2) + \epsilon(\vec{k}+\vec{K}_x/2)}{2} = \frac{\hbar^2}{2m}\big(k^2 + \frac{1}{4}K_x^2\big)-\mu\nn\\
\epsilon_2(\vec{k})&\equiv&\frac{\epsilon(\vec{k}-\vec{K}_x/2) - \epsilon(\vec{k}+\vec{K}_x/2)}{2} = -\frac{\hbar^2}{2m}\vec{k}\cdot\vec{K}_x\nn\\
D(\vec{k})&\equiv&\sqrt{\epsilon_2(\vec{k})^2 + U^2}
\eea
The eigenvectors and the original fermionic operators are related by
\bea
c_{\vec{k}-\vec{K}_x/2,\sigma}=\cos\theta_k c_{+,\vec{k}\sigma} - \sin\theta_k c_{-,\vec{k}\sigma}\nn\\
c_{\vec{k}+\vec{K}_x/2,\sigma}=\sin\theta_k c_{+,\vec{k}\sigma} + \cos\theta_k c_{-,\vec{k}\sigma},
\label{eigenvector}
\eea
where $\cos 2\theta_k = \frac{\epsilon_2(\vec{k})}{D(\vec{k})}$, and $0\leq \theta_k\leq \frac{\pi}{2}$.

It is important to note that instead of using quasimomentum on a tight-binding model in the reduced Brillouin-zone scheme, we have chosen to work on real momentum from a full Hamiltonian with 
periodic potential included explicitly. The advantage of our choice is that Umklapp processes in this formalism correspond to processes conserving the real momentum, but not the momentum 
in the band we are interested in. Therefore, Umklapp channels can be expressed as a series of density operators with matrix elements as functions of $\theta_k$, which can be done in 
a straightforward way. We will see how this works in the next section.

\section{RPA Theory for Density-Density Correlation Function in Normal State}
\begin{figure}
\includegraphics{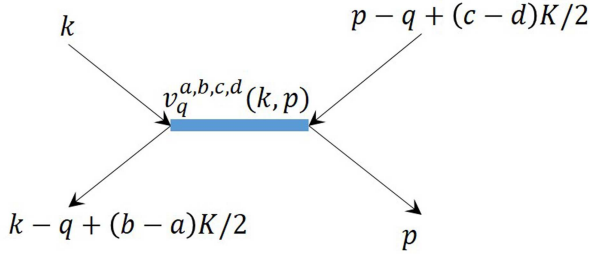}
\caption{\label{fig:vertex} Interaction vertex lines with Umklapp scattering. $V^{a,b,c,d}_q(\vec{k},\vec{p})$ is given in Eq. \ref{coul}.}
\end{figure}

In order to extract the Umklapp processes from the Hamiltonian in Eq. (\ref{hamil}), we need to expand $H_{Coul}$ in the band basis to determine the vertex lines required in the diagrammatic approach.
With the consideration of the periodic potential along the $\hat{x}$ direction, the annihilation operator in Eq. \ref{plane} can be written as
\be
\psi_{\vec{r},\sigma}= \sum_{a\in 2Z+1} \sum_{\vec{k}} e^{i(\vec{k}-a\vec{K}_x/2)\cdot\vec{r})}  c_{\vec{k}-a\vec{K}_x/2,\sigma}
\ee
By using Eq. (\ref{eigenvector}), we can express $c_{\vec{k}\pm\vec{K}_x/2,\sigma}$ in terms of Bloch bands $c_{\pm,\vec{k}\sigma}$. We restrict our interest only in the band on which the Fermi surface lies 
to highlight the features emerging entirely due to the Umklapp processes instead of interband scatterings. Assume that the Fermi surface lies on the $c_{-,\vec{k}\sigma}$ band, in which
the components in $H_{Coul}$ involving only $c_{-,\vec{k}\sigma}$ band are
\bea
&&H^\beta_{Coul}=
\frac{1}{2\Omega}\sum_{\vec{q}\neq 0} \sum_{a,b,c,d=-1,1}\sum_{\vec{k}\sigma} \sum_{\vec{p}\sigma'} V^{a,b,c,d}_q(\vec{k},\vec{p}) \nn\\
&&c^\dagger_{-,\vec{k}\sigma}c_{-,\vec{k}-\vec{q}+(b-a)\vec{K}_x/2\sigma}
c^\dagger_{-,\vec{p}-\vec{q}+(c-d)\vec{K}_x/2\sigma'}c_{-,\vec{p}\sigma'}
\label{coul}
\eea
where $V^{a,b,c,d}_q(\vec{k},\vec{p})$ can be read off Eq. (\ref{eigenvector}).
Now it is clear that Eq. (\ref{coul}) describes the components of the Coulomb interaction on $c_{-,\vec{k}\sigma}$ with both normal and Umklapp processes included, and the corresponding vertex lines 
are plotted in Fig. \ref{fig:vertex}.

The final form of $H_{Coul}$ in our consideration becomes
\be
H_{Coul} = \frac{1}{2\Omega}\sum_{\vec{q}\neq 0} v_q \hat{\rho}^{full}(\vec{q}) \hat{\rho}^{full}(-\vec{q})
\ee
where
\bea
\hat{\rho}^{full}(\vec{q}) &\equiv& 
\sum_{\vec{k},\sigma} \cos(\theta_{\vec{k}}-\theta_{\vec{k}-\vec{q}}) c^\dagger_{-,\vec{k}\sigma} c_{-,\vec{k}-\vec{q},\sigma}\nn\\
&-& \sum_{\vec{k},\sigma} \sin\theta_{\vec{k}}\cos\theta_{\vec{k}-\vec{q}}c^\dagger_{-,\vec{k}\sigma} c_{-,\vec{k}-\vec{q}-\vec{K},\sigma}\nn\\
&-& \sum_{\vec{k},\sigma} \cos\theta_{\vec{k}}\sin\theta_{\vec{k}-\vec{q}}c^\dagger_{-,\vec{k}\sigma} c_{-,\vec{k}-\vec{q}+\vec{K},\sigma}\nn\\
\eea
Now we can follow the standard approach to perform the generalized random-phase approximation (RPA) calculations. First we define
\bea
\hat{\rho}_1(\vec{q}) &\equiv& \sum_{\vec{k},\sigma} \cos(\theta_{\vec{k}}-\theta_{\vec{k}-\vec{q}}) c^\dagger_{-,\vec{k}\sigma} c_{-,\vec{k}-\vec{q},\sigma}\nn\\
\hat{\rho}_2(\vec{q}) &\equiv& -\sum_{\vec{k},\sigma} \sin\theta_{\vec{k}}\cos\theta_{\vec{k}-\vec{q}} c^\dagger_{-,\vec{k}\sigma} 
c_{-,\vec{k}-\vec{q}-\vec{K},\sigma}\nn\\
\hat{\rho}_3(\vec{q}) &\equiv& - \sum_{\vec{k},\sigma} \cos\theta_{\vec{k}}\sin\theta_{\vec{k}-\vec{q}}
c^\dagger_{-,\vec{k}\sigma} c_{-,\vec{k}-\vec{q}+\vec{K},\sigma}
\label{density}
\eea
The bare susceptibility becomes a matrix and each component can be expressed by the Lindhard function,
\bea
&&\hat{\chi}^0_{\hat{\rho}_i,\hat{\rho}_j}(\vec{q},\omega)\nn\\ 
&=& -\frac{2}{\Omega}\sum_{\vec{k},\sigma} f_if_j\times\big(\frac{n_F(E^-(k-q))-n_F(E^-(k))}{\hbar\omega + i\delta
+E^-(k-q) - E^-(k)}\big)\nn
\label{chi0rpa}
\eea
where
\bea
f_1(\vec{k},\vec{q}) &=& \cos(\theta_{\vec{k}}-\theta_{\vec{k}-\vec{q}}),\nn\\
f_2(\vec{k},\vec{q}) &=& - \sin\theta_{\vec{k}}\cos\theta_{\vec{k}-\vec{q}},\nn\\
f_3(\vec{k},\vec{q}) &=& - \cos\theta_{\vec{k}}\sin\theta_{\vec{k}-\vec{q}},
\label{f123}
\eea
and the factor of two in Eq. (\ref{chi0rpa}) comes from the spin degrees of freedom.
The final expression of the density-density response function with RPA is
\be
\big(\hat{\chi}(\vec{q},\omega)\big)^{-1} = \big(\hat{\chi}^0(\vec{q},\omega)\big)^{-1} + \hat{U}_q
\label{rpachi}
\ee
where the interaction kernel 
\be
\hat{U}_q=
\left(
\begin{array}{ccc}
v_q&v_{\vec{q}+\vec{K}}&v_{\vec{q}-\vec{K}}\\
v_{\vec{q}-\vec{K}}&v_q&v_{\vec{q}-2\vec{K}}\\
v_{\vec{q}+\vec{K}}&v_{\vec{q}+2\vec{K}}&v_q
\end{array}\right)
\ee
In the limit of long wavelength (small $q$), $f_1\approx 1$ and $v_{\vec{q}\pm\vec{K}}<<v_q$. As a result, the density-density response function in the normal process that we are interested 
in is $\left[\hat{\chi}(\vec{q},\omega)\right]_{1,1}$,
which describes the scatterings between particle-hole pairs with a total momentum of $\vec{q}$. $\left[\hat{\chi}(\vec{q},\omega)\right]_{2,2}$ and $\left[\hat{\chi}(\vec{q},\omega)\right]_{3,3}$ 
describe the scatterings between particle-hole pairs with a total momentum of $\vec{q}\pm K_x$. Off-diagonal terms in $\hat{\chi}(\vec{q},\omega)$ describe the 
scatterings between particle-hole pairs whose total momenta differ by $nK_x$, which are just the Umklapp processes.

\begin{figure}
\includegraphics{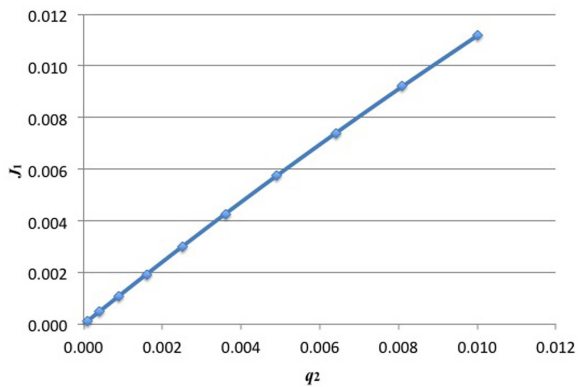}
\caption{\label{fig:fsum} Check of $f$-sum rule $J_1(\vec{q})$ as a function of $\vec{q}$. $J_1(\vec{q})$ scales with $q^2$ as expected.}
\end{figure}

It is worth mentioning that our formalism sattisfies the $f$-sum rule,
\be
J_1(\vec{q}) = \frac{2}{\pi}\int_0^\infty d\omega \omega\Im \big[\hat{\chi}(\vec{q},\omega)\big]_{11} = \frac{n q^2}{m},
\ee 
which only depends on $\vec{q}$ as the Fermi energy is fixed. 
We have checked that $J_1(\vec{q})$ is the same in both normal and superconducting states 
(the formalism for the superconducting state will be discussed in the next section), scaling with $q^2$ as shown in Fig. \ref{fig:fsum}. 

To demonstrate the features emerging from the Umklapp processes, we introduce an effective parameter $V_{Um}$ into $\hat{\chi}(\vec{q},\omega)$,
\bea
\left[\hat{\chi}'(\vec{q},\omega)\right]_{i,j} &=& \left[\hat{\chi}(\vec{q},\omega)\right]_{i,j}, i=j\nn\\
&=&  V_{Um}\left[\hat{\chi}(\vec{q},\omega)\right]_{i,j}, i\neq j.
\eea
It is instructive to analyze the case of $V_{Um}=0$ first. In this case, $\chi(\vec{q},\omega)$ only has the diagonal terms and a collective excitation occurs when 
\be
\big(\hat{\chi}^0(\vec{q},\omega)\big)_{ii}^{-1} + v_q = 0
\ee
The collective excitation in the $i=1$ channel is just the familiar plasmon excitation. We have checked that the frequency of this excitation scales with $\sqrt{q}$ as expected from a 
two-dimensional system. For $i=2,3$, there are collective excitations enabled entirely due to the periodic potential $U$. 
To see this, one can check that if $U=0$, then $f_2=f_3=0$ in Eq. (\ref{f123}). Consequently, no any collective excitation is present in $i=2,3$ channels. On the 
other hand, finite $U$ results in nonzero $f_2$ and $f_3$, producing collective excitations at energies lower than the plasmon excitation due to the fact that $f_2,f_3<1$. However, 
since there are no Umklapp processes (no off-diagonal terms), these collective excitations cannot be seen in $\left[\hat{\chi}(\vec{q},\omega)\right]_{11}$ channel, i.e., the density-density 
correlation function of experimental interest. 

\begin{figure}
\includegraphics{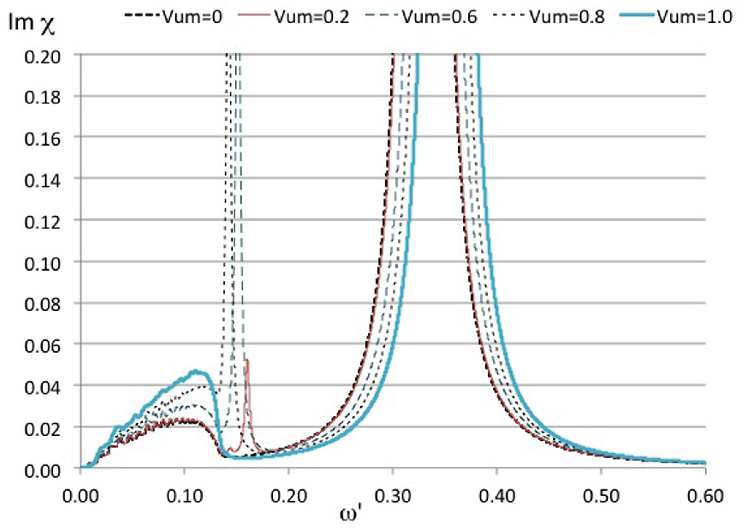}
\includegraphics{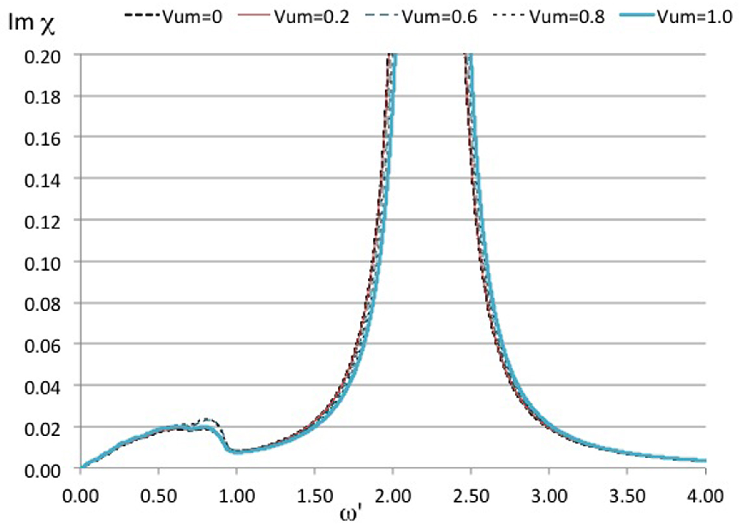}
\caption{\label{fig:imchi-vary-vum} (Upper) ${\rm Im} \chi(\vec{q},\omega)$ with $\mu=0.2$, $U=0.1$, $\alpha=0.1$, and $\vec{q}=(0.05,0)$ for different values of $V_{Um}$. The units of the
energy and momentum are chosen to be $\hbar^2 K_x^2/2m$ and $K_x$ respectively. In the plot, $\omega' = \hbar\omega/\mu$ and the strong delta-function peak around $\omega'=0.3$
corresponds to the plasmon excitation. As the strength of Umklapp scattering increases, the spectral weight
of the plasmon excitation transfers into lower frequency and the plasmon frequency increases. (Lower) The same plot for $\mu=0.02$ which gives a much smaller Fermi surface around the
$\Gamma$ point. In this case, the Fermi wavevector is much smaller than $\frac{K_x}{2}$, and consequently the effect of Umklapp scattering is negligible. Note that in both plots,
a broadening factor of $\gamma=0.0005$ is introduced.}
\end{figure}

Nevertheless, once the Umklapp processes are turned on, these collective excitations couple to the plasmon excitation, resulting in 
two important consequences. First, because the plasmon excitation is at highest energy, the couplings (Umklapp processes) push the plasmon frequency upward, while the collective excitations 
at $i=2,3$ channels are pushed downward. Second, because of the coupling, the collective excitations at the $i=2,3$ channels have finite spectral weights even in the 
$\left[\hat{\chi}(\vec{q},\omega)\right]_{11}$ channel. These features are clearly shown in Fig. \ref{fig:imchi-vary-vum}, which exhibits the increase of the plasmon frequency as well as the 
increase of the spectral weight at frequencies below the plasmon excitation with increasing $V_{Um}$. 
As the Umklapp scattering is strong enough so that the collective excitations at the $i=2,3$ channels are 
pushed into the particle continuum, a broad spectrum emerges.
The effects due to the Umklapp processes described above can be further checked by studying the case with small Fermi surface centered around the $\Gamma$ point. 
In this case, the Umklapp scattering should be strongly 
suppressed due to the energy conservation, which is verified in the lower plot of Fig. \ref{fig:imchi-vary-vum}.

The physics discussed above is very general. If we include more $l$'s in Eq. (\ref{hk}), the size of the $\hat{\chi}(\vec{q},\omega)$ matrix increases, producing more and more 
spectral weights at frequencies lower than the plasmon frequency. One can include periodic potential along the $\hat{y}$ direction as well, which corresponds to a two dimensional lattice system. 
The main effect of this is the increase of the size of the 
$\hat{\chi}(\vec{q},\omega)$ matrix, which produces more collective excitation at frequencies below the plasmon excitation and, consequently, more spectral weight transfers. 
Therefore, we conclude that with the inclusion of Umklapp scattering, 
significant amounts of spectral weights are transferred from the plasmon excitation to the lower energy, even for a single-band system. 
This is fundamentally different from the case without Umklapp scattering in which the plasmon mode is the only excitation at long wavelength and holds all the spectral weights.

\section{RPA Theory for Density-Density Correlation Function in Superconducting State}
To describe the superconducting state and its related collective excitations, we introduce the pairing interaction in this general form:
\be
H_{SC} = -V_p\sum_{\vec{k},\vec{k}'} g_k g_{k'} c^\dagger_{-,\vec{k}\uparrow}c^\dagger_{-,-\vec{k}\downarrow}c_{-,-\vec{k}'\downarrow}c_{-,\vec{k}'\uparrow}
\ee
where $g_k$ describes the gap symmetry which equals 1 for $s$-wave and $\frac{k_x^2 - k_y^2}{k^2_F}$ for $d$-wave superconductors.
Employing the mean-field theory on $H_{SC}$, we obtain the superconducting groundstate, and Bogoliubov quasiparticles ($\alpha$,$\beta$) read
\bea
c_{-,\vec{k}\uparrow}=\cos\phi_k \alpha_{\vec{k}} + \sin\phi_k \beta_{\vec{k}}\nn\\
c^\dagger_{-,-\vec{k}\downarrow}=-\sin\phi_k \alpha_{\vec{k}} + \cos\phi_k \beta_{\vec{k}},
\label{eigenvector-sc}
\eea
where $\phi_k = \frac{sgn(g_k)}{2} \cos^{-1}\big[\frac{E^-(\vec{k})}{E^{SC}(\vec{k})}\big]$, $E^{SC}(\vec{k}) = \sqrt{(E^-(\vec{k}))^2+(\Delta(\vec{k}))^2}$, and $\Delta(\vec{k}) = \Delta_0 g_k$. 
$\Delta_0$ is obtained by solving the gap equation of
\be
\frac{1}{V_p} = \sum_{\vec{k}} \frac{g^2_k}{2 E^{SC}(\vec{k})}.
\ee

Due to the nature of the Cooper pairs, the density-density correlation function is coupled to the pairing-pairing correlation function.
The pairing channel can be divided into phase $\Phi(\vec{q})$ and the amplitude $M(\vec{q})$ modes as
\bea
\Phi(\vec{q}) &=& \sum_{\vec{k}} g_k\big[c_{-,\vec{k}-\vec{q},\uparrow}c_{-,-\vec{k},\downarrow} - c^\dagger_{-,-\vec{k}+\vec{q}\downarrow}c^\dagger_{-,\vec{k}\uparrow} \big]\nn\\
M(\vec{q}) &=& \sum_{\vec{k}} g_k\big[c_{-,\vec{k}-\vec{q}\uparrow}c_{-,-\vec{k}\downarrow} + c^\dagger_{-,-\vec{k}+\vec{q},\downarrow}c^\dagger_{-,\vec{k}\uparrow} \big].
\eea
Together with $\rho_{1,2,3}(\vec{q})$ derived in the last section, now the susceptibility in the superconducting state is a $5\times 5$ matrix. We define
\be
A_{1,2,3}(\vec{q}) = \rho_{1,2,3}(\vec{q}), A_4(\vec{q}) = \Phi(\vec{q}), A_5(\vec{q}) = M(\vec{q}),
\ee
and the bare susceptibility with one-loop correction in the superconducting state is
\bea
&&\left[\hat{\chi}_{SC}^0(\vec{q},\omega)\right]_{A_i,A_j}\nn\\
&=& -\frac{1}{\Omega}\sum_{\vec{k},\sigma} \big(\frac{F_iF_j}{\hbar\omega + i\delta
-E^{SC}(k-q) - E^{SC}(k)}\nn\\
&-& \frac{G_iG_j}{\hbar\omega + i\delta + E^{SC}(k-q) + E^{SC}(k)}\big)\nn
\label{chisc0}
\eea
where
\bea
F_1&=&G_1 = f_1 \sin(\phi_{\vec{k}} + \phi_{\vec{k}-\vec{q}}),\nn\\
F_2&=&G_2 = f_2 \cos\phi_{\vec{k}}\sin\phi_{\vec{k}-\vec{q}} - f_3 \sin\phi_{\vec{k}}\cos\phi_{\vec{k}-\vec{q}},\nn\\
F_3&=&G_3 = f_3 \cos\phi_{\vec{k}}\sin\phi_{\vec{k}-\vec{q}} - f_2 \sin\phi_{\vec{k}}\cos\phi_{\vec{k}-\vec{q}},\nn\\
F_4&=& - G_4 = g_k \cos(\phi_{\vec{k}} - \phi_{\vec{k}-\vec{q}}),\nn\\
F_5&=&G_5 = g_k\cos(\phi_{\vec{k}} + \phi_{\vec{k}-\vec{q}}),\nn\\
\label{fg12345}
\eea
and $f_{1,2,3}$ can be found in Eq. \ref{f123}. The susceptibility at the RPA level in the superconducting state leads to
\be
\big(\hat{\chi}_{SC}(\vec{q},\omega)\big)^{-1} = \big(\hat{\chi}_{SC}^0(\vec{q},\omega)\big)^{-1} + \hat{U}'_q
\label{rpachisc}
\ee
with the interaction kernel of
\be
\hat{U}'_q=
\left(
\begin{array}{ccccc}
v_q&v_{\vec{q}+\vec{K}}&v_{\vec{q}-\vec{K}}&0&0\\
v_{\vec{q}-\vec{K}}&v_q&v_{\vec{q}-2\vec{K}}&0&0\\
v_{\vec{q}+\vec{K}}&v_{\vec{q}+2\vec{K}}&v_q&0&0\\
0&0&0&-\frac{V_p}{2}&0\\
0&0&0&0&-\frac{V_p}{2}
\end{array}\right)
\ee

\begin{figure}
\includegraphics{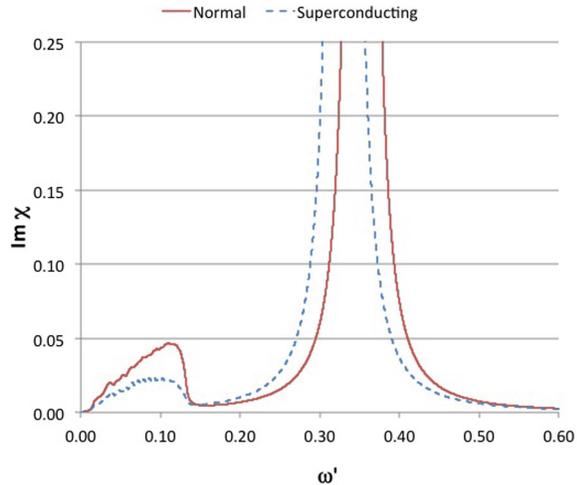}
\caption{\label{fig:imchi-sc} ${\rm Im} \chi(\vec{q},\omega)$ with $\mu=0.2$, $U=0.1$, $\alpha=0.1$, $V_{Um}=1.0$, and $\vec{q}=(0.05,0)$ in normal state and 
$d$-wave superconducting state with $\Delta_0 = 0.001$. The units and notation are the same as the ones used in Fig. \ref{fig:imchi-vary-vum}. The superconducting state suppresses the Umklapp scattering, 
resulting in the decreases of spectral weight below the plasmon excitation as well as the plasmon frequency.}
\end{figure}

It can be easily checked that if we turn off all the Umklapp processes by hand, only channels of $A_1$, $A_4$, $A_5$ are coupled to each others. In this case, 
we find that the plasmon excitation is still the only collective excitation in $\big(\hat{\chi}_{SC}(\vec{q},\omega)\big)_{11}$ and its frequency is the same as the frequency 
in the normal state. This is consistent with Anderson's theory\cite{anderson1958,anderson19582} as well as the sum-rule analysis done by Turkalov and Legget\cite{turlakov2003}. 
As the Umklapp scatterings are turned on, 
as shown in Fig. \ref{fig:imchi-sc}, we find that the effects of the Umklapp scatterings are much weaker in superconducting state than in the normal state.

To see how the superconductivity suppresses Umklapp processes, we analyze the crucial matrix elements of $ \hat{\chi}_{SC}^0(\vec{q},\omega)$ in Eq. (\ref{fg12345}).
Due to the pairing between electrons with $(\vec{k}\uparrow)$ and
$(-\vec{k}\downarrow)$ in the superconducting state, we need to rewrite the density operators in terms of Bogoliubov quasiparticles defined in Eq. \ref{eigenvector-sc}. 
Consequently, the density operators should be evaluated as follows
\bea
\hat{\rho}_i(\vec{q}) &=& \sum_{\vec{k},\sigma} f_i(\vec{k},\vec{q}) c^\dagger_{-,\vec{k}\sigma} c_{-,\vec{k}-\vec{q},\sigma}\nn\\
&=& \sum_{\vec{k}} \big[f_i(\vec{k},\vec{q}) c^\dagger_{-,\vec{k}\uparrow} c_{-,\vec{k}-\vec{q},\uparrow}\nn\\
&+& f_i(-\vec{k}+\vec{q},-\vec{q}) c^\dagger_{-,-(\vec{k}-\vec{q})\downarrow} c_{-,-\vec{k},\downarrow}\big].
\eea
From Eq. (\ref{f123}), we can easily see that $f_1(-\vec{k}+\vec{q},-\vec{q}) = f_1(\vec{k},\vec{q})$ while 
$f_2(-\vec{k}+\vec{q},-\vec{q}) = - f_3(\vec{k},\vec{q})$ and $f_3(-\vec{k}+\vec{q},-\vec{q}) = -f_2(\vec{k},\vec{q})$.
The crucial difference in the parity in the normal channel ($i=1$) and the 'Umklapp' channels ($i=2,3$) becomes important 
at small $q$. In this limit, $f_2\approx f_3$ so that in Eq. (\ref{fg12345}), $F_{2,3}\approx f_2\sin(\phi_{\vec{k}}-\phi_{\vec{k}-\vec{q}})\approx 0$. 
This indicates that the components involving 
$A_{2,3}$ channels are largely suppressed, and effectively only the $A_{1,4,5}$ channels dominate over the density-density correlation functions, resembling 
the case without Umklapp scattering. Therefore, the effects from the Umklapp
processes are largely suppressed by the superconductivity, and this conclusion is general for any gap symmetry. Physically, this suppression of Umklapp scattering is due to the 
interplay between the electron pairing and the odd parity of the matrix elements associated Umklapp processes, which can be seen directly from the above analysis on the $f_i(\vec{k},\vec{q})$.

The above analysis also shows that the superconductivity is particularly resistive to the Umklapp scatterings compared to other competing orders. Magnetic, charge-density wave, 
or nematic orders usually only induce the coupling between $\vec{k}$ and $\vec{k}+\vec{Q}$, where $\vec{Q}$ is the ordering wave vector, and, consequently, a large suppression due to 
the odd parity in $f_{2,3}(\vec{k},\vec{q})$ does not occur.

In summary, we predict a general feature for superconductivity emerging from a system with strong Umklapp scattering. The midinfrared spectrum in the density-density correlation function
decreases as the system has a phase transition from the normal to superconducting states, regardless of the gap symmetry. 
Meanwhile, the plasmon excitation has lower frequency and larger spectral weight in the superconducting state than in the normal state, which is consistent with the 
existing data of optical conductivity.
These consequences due to the suppression of 
Umklapp processes are unique in the superconducting state due to the interplay between the electron pairing and the odd parity of the Umklapp processes, which usually does not occur in other known 
competing orders.

\section{Comparison With Experiments}
\begin{figure}
\includegraphics{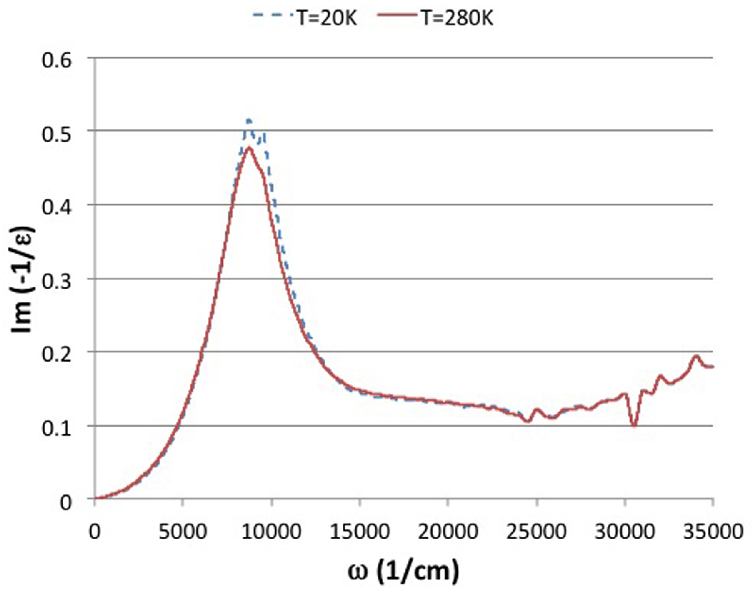}
\includegraphics{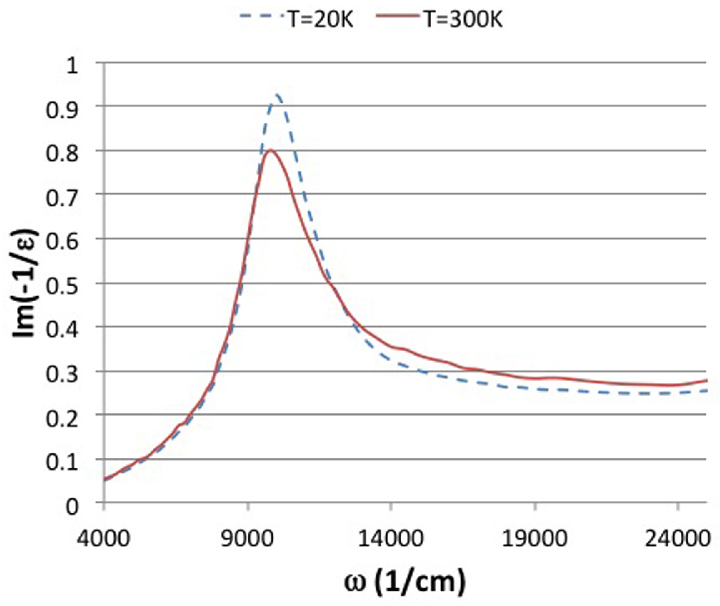}
\caption{\label{fig:loss} The loss function reconstructed from the original data of Ref. [\onlinecite{carbone2006}] (top) and the
recent measurement by Levallois {\it et. al.}\cite{julien} (bottom) on the optimally doped Bi$_2$Sr$_2$Ca$_2$Cu$_3$O$_{10}$ with $T_c = 110$ K. 
The pronounced peak around 9000 $cm^{-1}$ is the plasmon excitation.}
\end{figure}

As discussed in Ref. [\onlinecite{turlakov2003}], the electron energy loss spectroscopy (EELS)\cite{nucker1989,roth2010,schuster2012,kogar2014,roth2014} should be the most ideal probe 
for the density-density correlation function. 
The cross section of EELS
$\sigma(\vec{q},\omega)$ is typically interpreted as $\sigma(\vec{q},\omega) \propto v^2_q \Im \chi(\vec{q},\omega)$, where $\chi(\vec{q},\omega)$ is the true density-density correlation
function at $(\vec{q},\omega)$, which is $\big[\hat{\chi}(\vec{q},\omega)\big]_{11}$ in the present paper.
A systematic study of $\sigma(\vec{q},\omega)$ for cuprates with different dopings at different temperatures could be used to confirm the predction made above.

\begin{figure}
\includegraphics{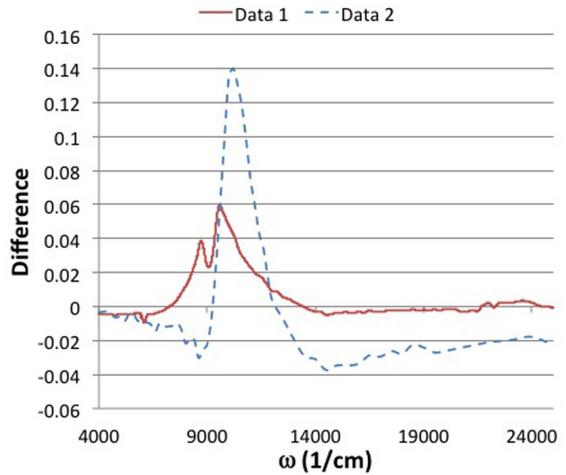}
\caption{\label{fig:diff} The difference in loss function between superconducting and normal states obtained from the original data of Ref. [\onlinecite{carbone2006}] (data 1) and 
recent measurement by Levallois {\it et. al.} (data 2) on the optimally doped Bi$_2$Sr$_2$Ca$_2$Cu$_3$O$_{10}$.}
\end{figure}

Relevant information can also be extracted from the existing data of optical conductivity.
It is well accepted that the $ab$-plane dielectric function can also be related to $\sigma(\vec{q},\omega)$ via
\be
\sigma(\vec{q},\omega)\propto \frac{1}{q^2}\Im\big[-\frac{1}{\epsilon_{ab}(\vec{q},\omega)}\big].
\ee
The quantity $\Im\big[-\frac{1}{\epsilon_{ab}(\vec{q},\omega)}\big]$, known as the loss function, is plotted in Fig. \ref{fig:loss} using the original data of Ref. [\onlinecite{carbone2006}] 
as well as the recent measurement by Levallois {\it et. al.}\cite{julien} on the optimally doped Bi$_2$Sr$_2$Ca$_2$Cu$_3$O$_{10}$.
The reconstructed loss function revealed two important features\cite{pm}.
First, pronounced peaks around 9000 $cm^{-1}$ ($\sim 1.1$ eV) could clearly be seen in the loss function. Second, the difference in the loss function between superconducting and normal states 
is plotted in Fig. \ref{fig:diff}. It exhibits an increase of the 
spectral weight of the pronounced peaks and a decrease of the spectral weight in a wide range of the lower frequencies as the system 
has a phase transition into the superconducting state. 
If the pronounced peaks around 1.1 eV are intepreted as the plasmon excitation, this observed change of spectral weight in the loss function in the frequency range below the plasmon excitation 
is qualitatively consistent with the present theory.
Early EELS data on cuprates\cite{nucker1989} obtained the plasmon energy to be $\sim 1$ eV. These results provide strong support for the present theory.

There exist, however, some features that cannot be descrbied by the presented theory. First, experimentally it has been found that spectral weight transfers also occur at an energy scale higher that 
the plasmon excitation, which is in the range of 1-3 eV. These transfers might result from the interband transitions that have been neglected in the present study. 
Second, because the plasmon peak is inevitably 
damped in the real materials, it is hard to judge whether or not the plasmon frequency decreases or not in the superconducting state, though its spectral weight clearly increases. 
Further study including interband
transitions as well as a broadening mechanism will be necessary to address these issues.

\section{Conclusion}

In this paper, we have investigated the effects of Umklapp processes on the density-density correlation function in both the normal and superconducting state. 
Without Umklapp processes, the plasmon mode is the only observable excitation at long wavelength in the density-density correlation function, and its energy and spectral weight are 
not affected by superconductivity at all, as shown in the seminal papers of Anderson\cite{anderson1958,anderson19582} on the theory of plasmon excitation in superconductors.
In the presence of Umklapp processes, we have shown that the plasmon mode is no longer the only excitation having finite spectral weight, and a significant amount 
of spectral weight is transferred from the plasmon excitation to lower frequency. 
In the superconducting state, the interplay between the nature of electron pairing between $(\vec{k}\uparrow)$ and $(-\vec{k}\downarrow)$  and the odd parity of the matrix elements associated 
with Umklapp processes substantially suppresses the effects from Umklapp processes. The important experimental signatures reflecting this superconductivity-induced suppression of 
Umklapp processes are the decrease of the spectral weights in the frequency range well above the gap but below 
the plasmon frequency, a downward shift of plasmon frequency, and an increase of plasmon spectral weight as the system undergoes the superconducting phase transition. 
Since all of the high-temperature superconductors known to date are crystalline, such effects should be
generally observable among these materials.
These signatures have been observed in the loss function of the optimally doped Bi$_2$Sr$_2$Ca$_2$Cu$_3$O$_{10}$, but further experiments on 
optical conductivity and EELS will be necessary.

\section{Acknowledgement}
We are grateful for valuable discussions with P. Abbamonte, A. Chubukov, W. Hanke,  J. Levallois, A. J. Leggett, D. Pouliot, P.W. Phillips, S. Raghu, D. J. Scalapino, and D. van der Marel.
We are particularly grateful to  J. Levallois and D. van der Marel for sharing the original data of Ref. [\onlinecite{carbone2006}] as well as the unpublished data on 
the optimally doped Bi$_2$Sr$_2$Ca$_2$Cu$_3$O$_{10}$.
W.-C.L. thanks KITP for the hospitality at UCSB while this paper was being finalized.
This work is supported by a start up fund from Binghamton University and in part by the NSF under Grant No. NSF PHY11-25915 for the KITP program 
'Magnetism, Bad Metals and Superconductivity: Iron Pnictides and Beyond'.

\end{document}